# Divergent effects of laser irradiation on ensembles of nitrogen-vacancy centers in bulk and nano-diamonds: implications for biosensing

Domingo Olivares-Postigo[1,2,3]*, Federico Gorrini[2,4], Valeria Bitonto[3], Johannes Ackermann[5], Rakshyakar Giri[1], Anke Krueger[5,6] and Angelo Bifone[2,3,4]

[1] Center for Neuroscience and Cognitive Systems, Istituto Italiano di Tecnologia, Corso Bettini 31, 38068 Rovereto, Trento, Italy
[2] University of Torino, Molecular Biology Center, via Nizza 52, 10126, Torino, Italy
[3] University of Torino, Department of Molecular Biotechnology and Health Sciences, via Nizza 52, 10126, Torino, Italy
[4] Istituto Italiano di Tecnologia, Center for Sustainable Future Technologies, via Livorno 60, 10144, Torino, Italy
[5] Institut für Organische Chemie, Julius-Maximilians-Universität Würzburg, Am Hubland, 97074 Würzburg, Germany
[6] Wilhelm Conrad Röntgen Center for Complex Materials Research (RCCM), Julius-Maximilians University Würzburg 97074, Germany

## Abstract

Ensembles of negatively charged nitrogen vacancy centers ($NV^-$) in diamond have been proposed for sensing of magnetic fields and paramagnetic agents, and as a source of spin-order for the hyperpolarization of nuclei in magnetic resonance applications. To this end, strongly fluorescent nanodiamonds represent promising materials, with large surface areas and dense ensembles of $NV^-$. However, surface effects tend to favor the less useful neutral form, the $NV^0$ centers. Here, we study the fluorescence properties and optically detected magnetic resonance (ODMR) of $NV^-$ centers as a function of laser power in strongly fluorescent bulk diamond and in nanodiamonds obtained by nanomilling the native material. In bulk diamond, we find that increasing laser power increases ODMR contrast, consistent with a power-dependent increase in spin-polarization. Surprisingly, in nanodiamonds we observe a non-monotonic behavior, with a decrease in ODMR contrast at higher laser power that can be ascribed to more efficient $NV^-\rightarrow NV^0$ photoconversion in nanodiamonds compared to bulk diamond, resulting in depletion of the $NV^-$ pool. We also studied this phenomenon in cell cultures following internalization of NDs in macrophages. Our findings show that surface effects in nanodiamonds substantially affect the NV properties and provide indications for the adjustment of experimental parameters.

Keywords: nanodiamonds, polarization, nitrogen-vacancy centers, bulk diamond, spin dynamics, charge dynamics, charge stability, photoconversion, nanomilling, $^{13}C$, cells

* domingo.olivares@iit.it

# 1. Introduction

Negatively charged Nitrogen-Vacancy centers (NV$^-$) can be used as probes for ultrasensitive detection of magnetic [1–3] and electric fields [4–6], temperature [7,8], and nuclear spins at the nanoscale [9,10]. NV$^-$ centers can be optically polarized with continuous irradiation of laser light at room temperature and at Earth's magnetic field [11–13], thus paving the way for applications in biomedical assays [14–17], as intracellular thermometers [7,18,19], optical magnetic imaging in living cells [2,20,21] or optical magnetic detection of single-neuron action potentials [22,23]. Moreover, optical pumping of NV$^-$ centers has been proposed as an alternative to dynamic nuclear polarization (DNP) for the hyperpolarization of nuclear spins [24–26].

These applications rely on ensembles of NV$^-$ centers in the vicinity of the diamond surface [27,28]. Due to their large surface area, NV-rich, fluorescent nanodiamonds (NDs) are promising candidates for such applications [7,15,21]. Moreover, NDs are biocompatible, and their surface can be functionalized to target specific cells or proteins [29–31]. NDs can be internalized by living cells, thus probing the microenvironment in subcellular compartments [32,33]. Furthermore, $^{13}$C enrichment of NDs is desirable to improve the hyperpolarization efficiency for contrast agents, quantum sensing and magnetic resonance signal enhancement [34].

Several techniques have been proposed for the production of NDs, including nanomilling [35] of bulk diamond, detonation diamond [36,37] and high-power laser ablation [38,39]. Detonation NDs tend to be small (5-10 nm) and rich in impurities [36]. High-power laser ablation can produce fluorescent NDs in a single-step process [38,39], but the yield is insufficient for practical application. Conversely, nanomilling of bulk diamond makes it possible to control concentration of defects and NV centers, as well as particle size, with good production yield [40].

Increasing the surface area of the material by nanomilling can improve exposure of shallow NV centers to the external environment. However, it inevitably affects NV$^-$ charge stability [41], as surface effects tend to favor the NV$^0$ charge state, which does not present useful spin properties [11–13]. As a result, NDs present larger relative concentration of NV$^0$ in NDs compared to the starting material.

Laser light used to polarize and interrogate NV$^-$ centers can also induce charge switching between NV$^-$ and its neutral form NV$^0$ [27,42–47]. Photoconversion depends on the presence of nitrogen defects and surface acceptor states, and thus both NV$^-$ → NV$^0$ and NV$^0$ → NV$^-$ photoconversion routes have been observed in different diamond samples. Recently, we have shown that increasing laser power can substantially increase the availability of shallow NV$^-$ produced by nitrogen implantation in ultrapure CVD diamond [48]. However, it is unclear whether a similar strategy may be advantageous in NDs.

Here, we prepared NDs of various sizes by nanomilling of highly-fluorescent $^{13}$C-enriched diamond. Fluorescence spectra and optically detected magnetic resonance (ODMR) spectra were acquired at different laser powers for the native bulk diamond, and for NDs of 156 nm and 48 nm. Additionally, we internalized NDs in macrophage cells to study the effects of laser power under the typical conditions of a bioassay and in the cellular environment. Experiments were performed with a house-built wide-field microscope at sub-micrometric spatial resolution (1 pixel in the image equals 160 nm) to account for the intrinsic heterogeneity in NDs behavior and to study the microenvironment in different cellular compartments. Our findings highlight the importance of surface effects on photoconversion and provide useful information on the optimization of experimental conditions for biosensing or polarization transfer applications involving fluorescent NDs.

## 2.    Materials and Methods

**Samples**

Bulk $^{13}$C-enriched diamond grown by the high-pressure, high-temperature (HPHT) technique was acquired from ElementSix. These samples have a concentration of NVs of ≈10 ppm, and $^{13}$C enrichment ranged from 5 to 10% (≈$5*10^4$ - $10^5$ ppm), depending on the position within the diamond stone; concentration of substitutional nitrogen (P1 centers) was approximately 200 ppm. The fraction of $^{13}$C was provided by the vendor, while the concentrations of nitrogen and NV centers were estimated through spectroscopic measurements on the bulk diamond before milling [27]. Attrition milling was used to prepare samples with varying size distribution with a procedure adapted from [49]. The whole process is exemplified in Fig. 1a. The sample material (62 mg) was added to a stainless steel milling cup, which was filled up to one third of its height with 5 mm stainless steel milling balls. The bottom third of the milling cup was then filled with isopropanol that had been dried using molecular sieve. The sample was milled for six hours at 50 swings per second. After milling, the sample contained a significant amount of metallic debris caused by the milling. To clean the diamond material, the sample was flushed out of the milling cup with distilled water into a 250 ml round bottom flask. The remaining steel balls were removed using a magnet. To dissolve metallic impurities, 50 ml of concentrated hydrochloric acid was added. The mixture was stirred overnight at room temperature. After settling, the supernatant was decanted and 50 ml 96% of sulfuric acid was added. The resulting mixture was heated to 120

°C bath temperature without a reflux condenser to remove any remaining isopropanol. After one hour, a reflux condenser was attached to the flask and 20 ml of nitric acid (65%) slowly added while monitoring the reaction mixture carefully to prevent a violent reaction. It is important to note that any remaining isopropanol might violently react with concentrated nitric acid, thus it is important to remove it carefully before adding $HNO_3$! The solution was stirred overnight at 120 °C. After the solution had cooled down to room temperature, the supernatant acid mixture was removed using a pipette and the solution was transferred to centrifugation tubes. To wash the diamond material, centrifugation at 15000 rpm was used over one hour to settle the diamond material at the bottom. The supernatant was removed and replaced with distilled water. The diamond material was then dispersed using sonication. This process was repeated until the supernatant showed a neutral pH. After reaching neutral pH, centrifugation was used to separate the particles by size. Two fractions were separated, and their size distribution measured using dynamic light scattering, giving a D50 value (volume distribution) of 156 nm and 48 nm, respectively. NDs were suspended in deionized water and stored in glass vials. Data of DLS and scanning electron microscopy can be found in the Supplementary Material.

Prior to the PL and ODMR experiments, suspensions of NDs were sonicated, and 10 µL were deposited either on a circular glass slide or in ibides (Ibidi GmbH, Planegg/Martinsried, Germany) and allowed to dry. In a separate set of experiments, NDs were internalized into cells following the procedure described below.

**Cell line**

Murine (RAW 264.7) cell line was purchased from American Type Culture Collection (ATCC LGC Standards, Sesto San Giovanni, Italy) and cultured in DMEM supplemented with 10% (v/v) of FBS, 2 mM L-glutamine, 100 U/mL penicillin and 100 µg/mL streptomycin at 37 °C in a humidified atmosphere with 5% $CO_2$.

**In vitro NDs uptake experiments**

For uptake experiments, RAW 264.7 cells were seeded in an Ibidi at a density of $3 \times 10^4$ cells/ well and incubated at 37° C for 24 h, to allow them to adhere to the slide surface. Incubation of cells with NDs (size 156 nm and 48 nm) was performed for 24 h at 37 °C in a humidified atmosphere with 5% $CO_2$. At the end of the incubation, cells were washed three times with PBS and fixed in 4% PAF at room temperature for fifteen minutes.

**Confocal microscopy**

Cells were rinsed twice with phosphate buffered saline (PBS) and permeabilized with 0.1 % Triton in PBS for ten minutes. Actin filaments were stained with phalloidin fluorescein isothiocyanate (FITC) (Sigma) for thirty minutes at room temperature. After washing twice with PBS, nuclei were counterstained with 4',6-diamidino-2-phenylindole (DAPI). Coverslips were mounted with a glycerol/water solution (1/1, v/v). Observations were conducted under a confocal microscopy (Leica TCS SP5 imaging system) equipped with an argon ion and a 561 nm DPSS laser. Cells were imaged using a HCX PL APO 63×/1.4 NA oil immersion objective. NDs were excited by 561 nm laser, while the emission was collected in the 570–760 nm spectral range. Phalloidin was imaged using 458 nm laser and the emission collected in the 498-560 nm range. DAPI was imaged using 405 nm laser and the emission was collected in the 415-498 nm range. Image analysis was performed using ImageJ software.

**FL and ODMR experimental setup**

Full fluorescence spectra of the NVs were acquired with a confocal microRaman setup (LabRam Aramis, Jobin-Yvon Horiba), equipped with a DPSS laser (532 nm) and an air-cooled multichannel CCD detector in the window range between 535 nm and 935 nm.

Wide-field ODMR imaging was performed with a modified Nikon Ti-E inverted wide-field microscope [Fig. 1b,c] equipped with a microwave channel and a high-sensitivity CMOS camera (Hamamatsu ORCA-Flash4.0 V2) [39]. A 532 nm continuous-wave laser (Model: CNI laser mod. MGL-III-532/50mW) was used as excitation source, delivering on the sample a power of ~30 mW through a 40X (NA=0.75 and working distance of 0.66 mm) refractive objective. The laser power was modulated by inserting neutral density filters on the optical path. To avoid backscattering of laser light, we used a custom made dichroic beamsplitter. FL was collected in a spectral window ranging from 590 to 800 nm.

The microwave (MW) field source was a WindFreak MW generator (SynthHD v1.4 54MHz-13.6GHz); the signal was amplified by a Mini-Circuits ZVE-3W-83+ 2W RF amplifier. The MW irradiation was delivered to the sample with a MW single loop terminated with a high-power MW damper.

The temporal sequence of the experiment (camera acquisition and RF delivery) was controlled by a SpinCore 100 MHz TTL generator (Model: TTL: PB12-100-4K). Finally, the image acquisition was processed with the Nikon NIS-Elements Advanced Research software and analyzed with Fiji software.

The home-built instrument used for these experiments makes it possible to extract ODMR spectra pixelwise with sub-micrometric spatial resolution, thus enabling analysis of heterogeneously distributed samples. Indeed, the deposition procedure can result in a non-uniform distribution of NDs, with region-dependent concentration and size of aggregates. NDs uptake from cells is also inhomogeneous, and cells with varying amount and clustering of NDs inside were observed.

**ODMR technique description**

This modified wide-field microscope was used to perform spatially-resolved Continuous Wave ODMR (CW-ODMR), a technique to determine the sublevel structure of the ground state. The ground state of the NV center is a spin triplet, with the $|g, m_s = \pm 1\rangle$ state levels upshifted by 2.87 GHz with respect to the $|g, m_s = 0\rangle$ state. Continuous irradiation with a 532 nm laser polarizes the $|g, m_s = 0\rangle$ ground state level through spin dependent transitions from the excited state through the metastable singlet states. Different laser power levels in the range 1-30 mW are used for this experiment. MW frequency is swept in the 2.75-3.00 GHz region at a fixed power output of 15 dBm to detect $^{13}$C sidebands and NV$^-$ central resonances. In fact, when the MWs are resonant with the $|g, m_s = 0\rangle \leftrightarrow |g, m_s = \pm 1\rangle$ ground state transition, the darker states $|g, m_s = \pm 1\rangle$ become populated, and a drop in the fluorescence (a "dip" in the ODMR spectra) is recorded. The ODMR contrast provides a measure of spin polarization of the $|g, m_s = 0\rangle$. However, the ODMR spectrum is also affected by the relative contribution from NV$^0$ fluorescence, which is not modulated by MW and offsets background fluorescence. An increase in ODMR contrast may thus indicate increased polarization of NV$^-$, or decreased concentration of NV$^0$s. The sequence of laser and MW irradiation to perform the ODMR experiments is described in the box of Fig. 1b.

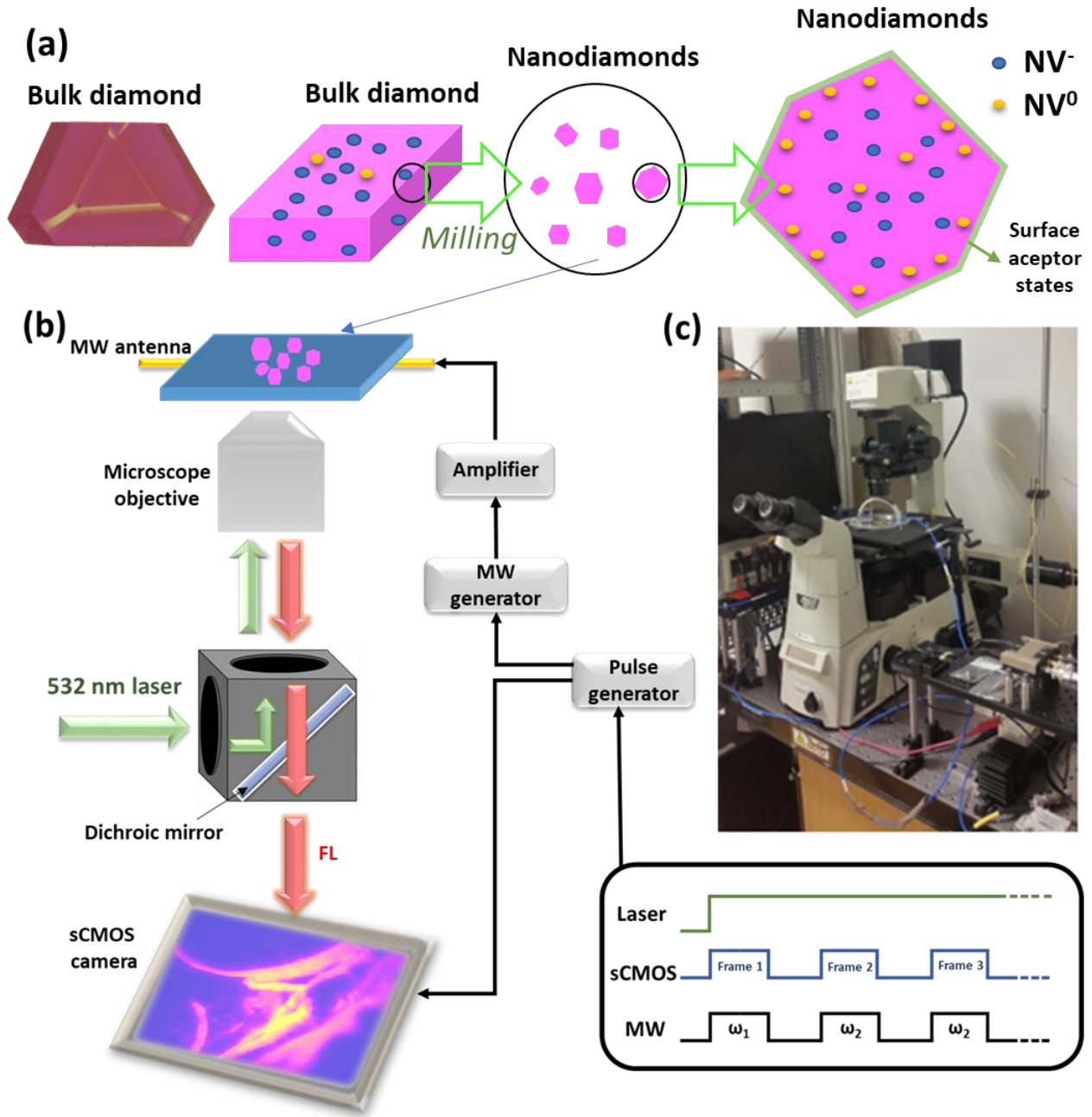

***Figure 1| Preparation of NDs and schematic of the wide-field ODMR set-up.*** *a) NDs, 156 nm and 48 nm in size were obtained by milling bulk, $^{13}$C-enriched HPHT fluorescent diamond. b) Schematics of the inverted wide-field ODMR microscope. Green laser light is reflected onto the sample. The fluorescence is collected by the objective through a dichroic mirror to the high-sensitivity CMOS camera. The image shows an actual fluorescence image of NDs deposited on a glass slide. MW delivery is obtained with a loop antenna. The panel shows the temporal diagram of the CW-ODMR experiment. The laser is kept ON during the entire measurement, while the camera detects the fluorescence image synchronously with the MW irradiation. c) Photo of the Wide-field ODMR microscope set-up.*

# 3. Results and discussion

Fig. 2 a,b,c show the fluorescence (FL) spectra from the bulk diamond, the 156 nm and the 48 nm NDs. Each NV charge state is characterized by its characteristic fluorescence spectrum, with zero-phonon lines (ZPL) at 575 nm and 638 nm for the $NV^0$ and $NV^-$, respectively. In addition, a phonon sideband, peaked around 620 nm for the $NV^0$ and 700 nm for the $NV^-$, is observed, extending up to ≈850 nm in both cases. On average, the two NDs samples presented a lower overall FL than bulk diamond (not apparent in figure 2 a,b,c, where the spectra are normalized to the $NV^-$ ZPL for comparison). Moreover, NDs showed a larger component from the $NV^0$s with respect to the bulk diamond (represented as black and light red curves, respectively). This is consistent with the idea that surface effects favor the $NV^0$ centers, thus affecting the relative concentration of the different charge states.

Fig 2 d,e,f show the ODMR spectra of the bulk diamond and NDs at different laser powers (from 1 to 30.5 mW) at constant MW power of 15 dBm. The ODMR spectra show the $NV^-$ central lines ($A_L$ and $A_R$) and two side bands ($B_L$ and $B_R$), the latter related to the hyperfine interaction between the $NV^-$ spins and the $^{13}C$ nuclear spins [50] . The $^{13}C$ sidebands are separated from each other by ~130 MHz, consistently with previously reported values [34,50–53]. For all samples, the linewidth of the resonance bands decreases with laser power, in agreement with the line-narrowing effect described by Jensen et al. [54]. Alongside with the increase in signal-to-noise (SNR) ratio, this phenomenon improves resolution of the $NV^-$ strain-split doublet (central dips $A_L$ and $A_R$). We did not observe any detectable temperature effect on the position of the ODMR resonance that may be caused by absorption of MW or laser power [4].

In bulk diamond, a monotonic increase in ODMR contrast with laser power was observed, consistent with increasing $NV^-$ polarization levels [Fig. 2d and 3a]. Contrast was calculated as the mean value from the central resonances $A_L$ and $A_R$. Interestingly, in NDs, a non-monotonic behavior was observed, as shown in Fig. 3 b,c. Up to 8.1 mW of laser power the ODMR contrast increases. At the highest power levels, the trend is the opposite, with a decrease in contrast systematically observed in different sample regions and for both ND sizes [Fig.3 b,c].

The inversion in the dependence of ODMR contrast in NDs appears paradoxical, as it would indicate a reduction in $NV^-$ spin-polarization with increasing laser power. However, this phenomenon is likely to reflect the different contributions of $NV^0$ centers in the bulk and ND samples under the various experimental conditions explored here. In fact, we notice that both NV- and $NV^0$ signals are collected by the broad bandpass filter with a spectral window of 590-800 nm (see Methods). The ODMR contrast is defined as $C_s = (I_{off} - I_{on})/I_{off}$, where $I_{off}$ and $I_{on}$ are the $NV^-$ FL intensities with MWs off- and on- resonance, respectively. However, the FL contains a contribution ($I_0$) from the NV$^0$ centers, which is not modulated by MWs and reduces the contrast by a factor $I_{off}/(I_{off} + I_0)$. This reduction factor contrast is very different for bulk and NDs. In the bulk, the vast majority of the NV centers are negative, as shown in Fig. 2a, and remain stable under laser irradiation. Therefore, a stronger laser irradiation results in a better spin initialization and improved ODMR contrast, without impacting on the NV charges (I$_0$ negligible). On the contrary, NDs tend to have higher relative concentrations of NV$^0$ centers as a result of surface effects (Fig. 2 a,b,c). Moreover, under laser irradiation, NV$^-$→ NV$^0$ photoconversion might be more efficient in NDs due to presence of surface acceptor states that can take a photoexcited electron from NV$^-$. Higher laser powers then result in increasing relative concentrations of NV$^0$ in NDs, and in a reduction in ODMR contrast. This phenomenon competes with NV$^-$ polarization, which increases with laser power, resulting in the non-monotonic behavior shown in Fig. 3 b,c. Therefore, in NDs there is an optimal laser power that should be used to prepare the NV$^-$ spin states. Stabilization of the NV$^-$ might shift the maximum of curve of Fig. 3 b,c to higher laser power.

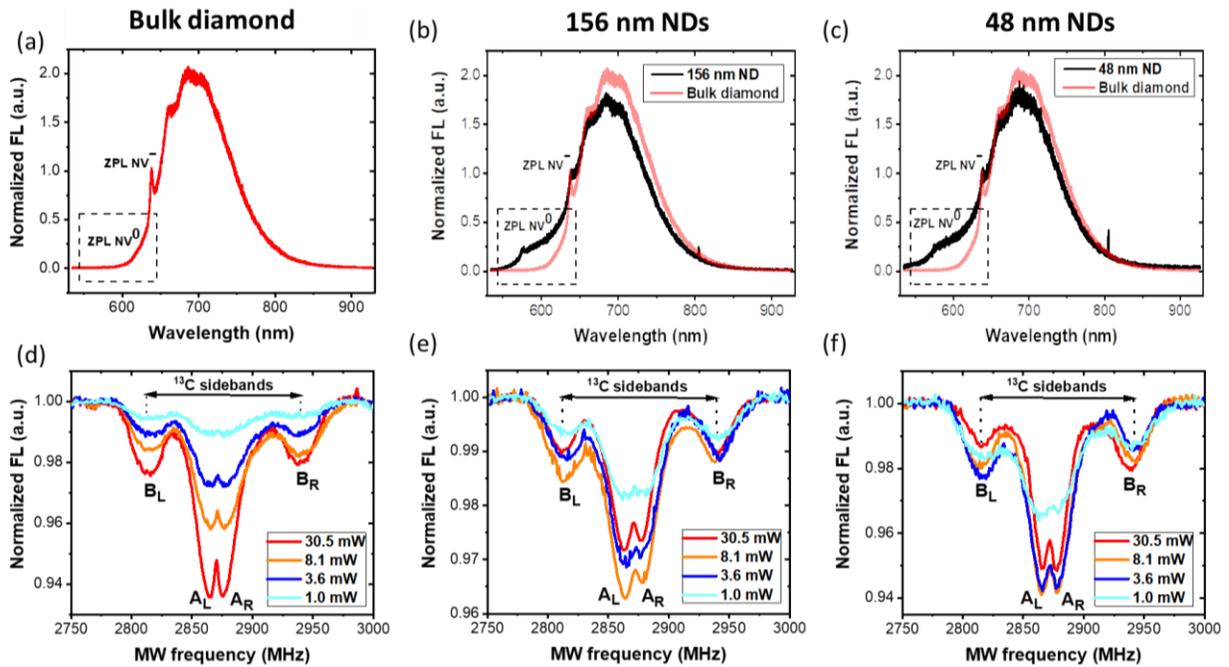

*Figure 2 | Fluorescence and ODMR spectra. Fluorescence and ODMR spectra for the (a,d) bulk diamond, (b,e) 156 nm and (c,f) 59 nm NDs, respectively. (a,b,c) NV$^0$ and NV$^-$ fluorescence spectra are characterized by ZPL lines at 575 nm (NV$^0$) and 638 nm (NV$^-$). Due to surface effects, NDs have a larger NV$^0$ component than bulk diamond, where the NV$^0$ band is practically undetectable. Panels (d,e,f) show ODMR spectra at different laser power with fixed MW power. In the ODMR spectra, the NV$^-$ central lines (A$_L$ and A$_R$) provide a measure of the spin polarization of the $|g, m_s = 0\rangle$ ground state. The MW range was set to 2.75-3 GHz to show the NV$^-$ central lines and the $^{13}$C sidebands (B$_L$ and B$_R$). In bulk diamond, ODMR contrast increases*

*with increasing laser power. Conversely, in the NDs, the ODMR contrast increases up to 8.1 mW and then, decreases at higher laser powers.*

A practical difficulty in assessing the properties of ensembles of NDs is the large variability of the optical response in heterogeneous samples [41,55–57]. This has been ascribed to inter-aggregate interactions, size distribution and different efficiency in NV center initialization in ND aggregates [57]. To circumvent this problem, we resorted to use a wide-field fluorescence microscope to spatially resolve ODMR spectra in different parts of the sample. The heterogeneity in ND deposition is apparent in Fig.3 b,c, where FL images clearly show inhomogeneous aggregation and concentration of NDs. ODMR contrast in these samples depends on the selected ROIs, characterized by different levels of FL [Inset Fig. 3], with lower variability observed in the 156 nm NDs compared to the 48 nm NDs. Conversely, the ODMR contrast from the bulk diamond is uniform throughout the image. Despite region and sample dependent contrast, our data show consistent trends for NDs in different samples and ROIs, thus suggesting that the effect is robust and reproducible. Charge dynamics and spin properties of NVs also depend on the ND microenvironment. To explore the effects described above in a typical bioassay, we incubated the NDs in cell cultures of macrophages (RAW 264.7). Internalization in macrophages is described in the Methods and illustrated in Fig. 4. The composite figures on the right show the NDs (orange) internalized in the cells, together with the actine filaments (green) and the nuclei (blue), for both 156 nm NDs (top row) and 48 nm NDs (bottom row) The concentration of the 156 nm NDs in liquid is much higher than that of the 48 nm NDs, resulting in a greater internalization of the larger NDs. Thus, only the results of 156 nm NDs are reported in the following section.

Fluorescence images were taken to evaluate the different amount of internalization of NDs in different cells. To this end, we extracted the ODMR contrast from ROIs of different size and location [Fig. 5a]. Representative examples of ODMRs curves extracted from single-cell ROIs (~6x6 μm) or a cell-aggregates ROI (from ~40x40 μm to ~100x100 μm) are shown in Fig. 5b and Fig. 5c, respectively. Despite some line broadening, compared to the bare NDs, the $NV^-$ central bands and $^{13}C$ sidebands can still be resolved. As for the previous samples, when increasing the laser power, the linewidth of the resonances decreases with the laser power, while the SNR increases, therefore improving the resolution of the $NV^-$ strain split doublet. Also in this case, we do not observe any variation depending on the temperature (i.e., no shift of central resonances), despite a possible MW absorption from the water in the cells or in the biological environment.

Fig. 5d shows the evolution of ODMR contrast with laser power for different ROIs. While qualitatively similar, the ODMR contrast dependence on laser power is more variable than the one observed in the bare NDs of Fig. 3b, ranging from a small reduction at the highest laser power, to a plateau-like behavior or even a slight increase. We speculate that this wider heterogeneity may be due to differences in the microenvironment, and particularly to the different pH of various cellular compartments (e.g., pH≈5 in lysosomes compared to pH≈7.2 in the cytoplasm).

Indeed, pH can affect the functional groups at the ND oxidized surface (carboxylic acids, ketones, alcohols, or esters), thus changing the properties and charge stability of shallow NV centers. At low pH, e.g., carboxylates will be protonated to a much higher extent than under physiological conditions at pH≈7.2, with potential effects on charge state of nearby NV centers. While plausible, this hypothesis requires further investigation.

We also studied the heterogeneity of the optical properties of NDs in cell in greater detail (Figure 5 e,f), with the help of a simple procedure. Here we acquired two FL images, with MW on- and off-resonance. Taking their difference and then normalizing pixelwise, it is possible to reconstruct an ODMR contrast image (Fig. 5g). In this ODMR contrast mapping, an average 4.5% contrast is observed. Moreover, parts with a high (red) and a low (blue) ODMR contrast can be imaged within the cell. These regions of low ODMR contrast correspond to cellular compartments where NDs cannot easily access, such as the nuclei. We selected a 1% contrast threshold in the ODMR mapping to cut down the noise. Therefore, wide-field ODMR mapping demonstrates the detection of NV centers in cells with a subcellular spatial resolution that is important for mapping the internalization of NDs in cellular compartments. These results pave the way for real-time, fast, and non-invasive mapping of single and ensemble of NV-enriched NDs in vivo and in vitro cellular environments.

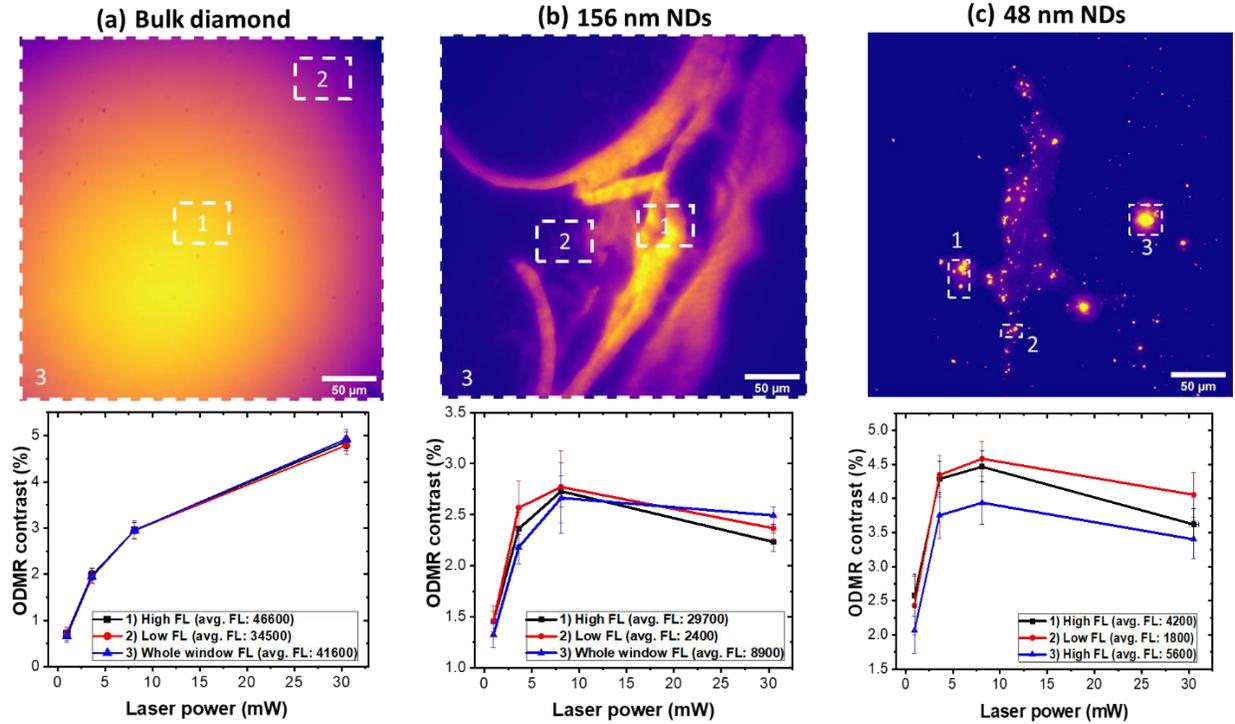

*Figure 3 | Comparison of ODMR contrast at different laser powers for bulk diamond and NDs. ODMR curves are extracted from three different regions of interest (ROIs) containing the bright spots that indicate presence of NV centers (see wide-field images). As laser power increases, in bulk diamond the ODMR contrast increases, consistently with larger $NV^-$ polarization levels (a). Conversely, in NDs of both sizes, the ODMR contrast increases up to 8.1 mW and decreases at higher laser powers, resulting in a non-monotonic behavior (b,c). The homogeneity of bulk diamond is reflected in the uniform values of ODMR contrast observed in the various ROIs with different levels of FL (inset). On the contrary, NDs show non-uniform deposition and aggregation, resulting in a region-dependent ODMR contrast. 156 nm NDs were less inhomogeneous than the 48 nm NDs. The power dependence of the ODMR signal was similar in all the ROIs extracted, for both NDs sizes.*

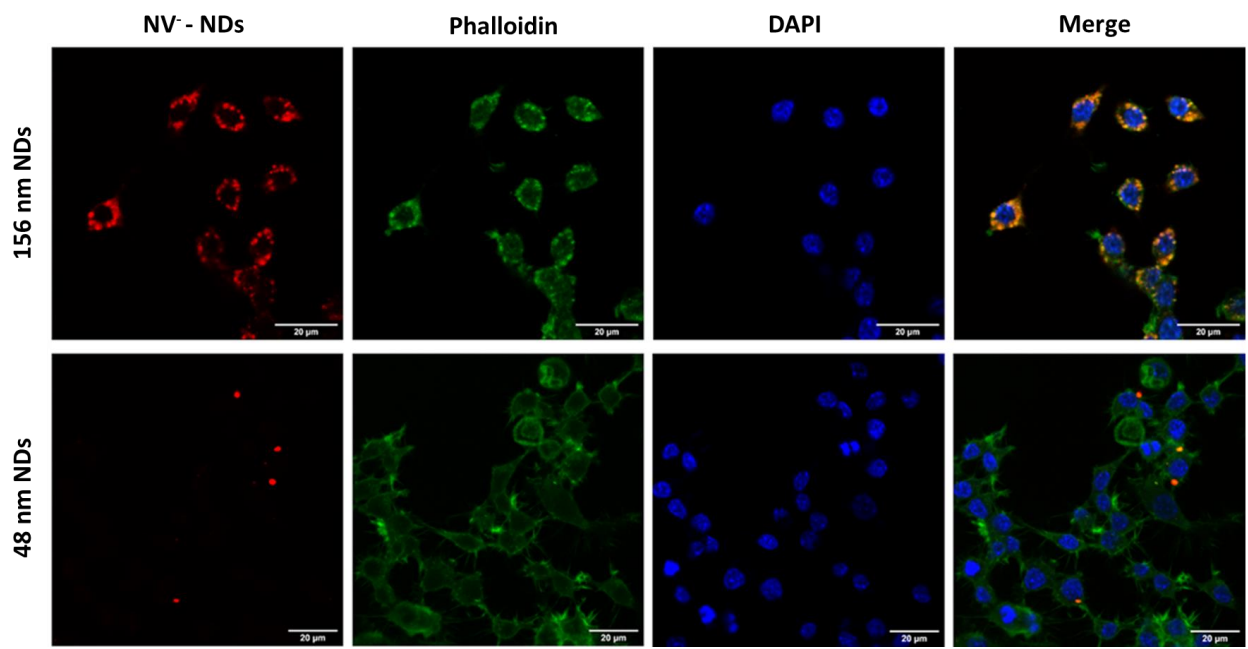

*Figure 4 | Confocal Microscopy images of RAW 264.7 cells incubated for 24 h with 48nm and 156nm NDs (red) and stained with Phalloidin (green). Nuclei were counterstained with DAPI (blue). The "merge" figure demonstrates the internalization of the NDs (orange) within the cells. Cellular uptake of the 156 nm NDs is much higher than that of the 48 nm NDs. Scale bars = 20μm. Representative images are shown.*

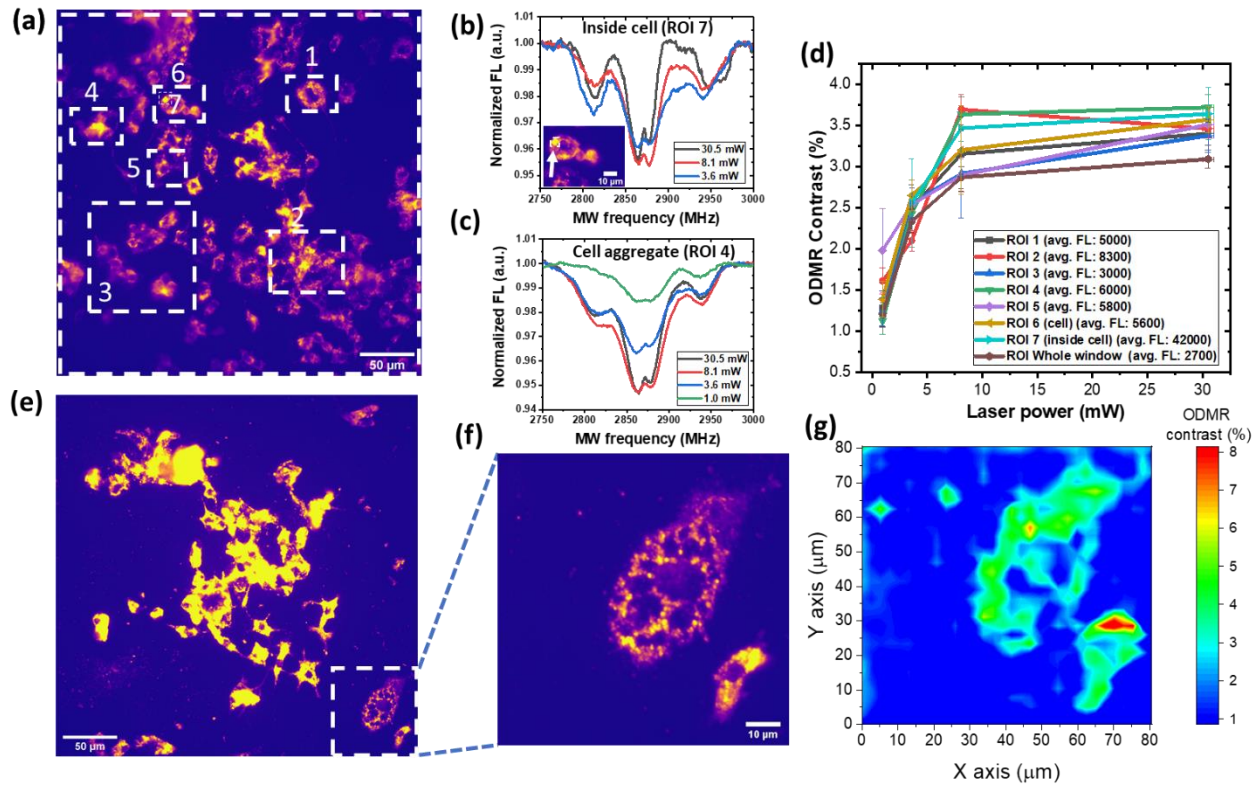

*Figure 5 | ODMR contrast comparison of NDs internalized in cells.* (a) FL image of 156 nm NDs internalized in macrophage cells; the white square delineates the ROIs from which signals were extracted. ODMR spectra showing the NV- central lines ($A_L$ and $A_R$) and the $^{13}C$ sidebands ($B_L$ and $B_R$) at different laser power with fixed MW power inside a cell are shown in (b), and in (c) for a cellular aggregate. (d) ODMR curves from different ROIs of various levels of fluorescence intensity. ODMR contrast steadily increases to 8.1 mW, while at higher laser powers it shows a variety of behaviors. NDs internalized within cells also show inhomogeneous aggregation and concentration, resulting in a region-dependent different ODMR contrast. (e) FL image of 156 nm NDs internalized in cells. The selected ROI (f) indicates a cell with its different compartments and its nucleus. (g) map of the distribution of ODMR contrast within the cells. Contrast varies regionally, with an average value of 4.5%.

# 4.   Conclusion

To summarize, we have studied fluorescence and optically detected magnetic resonance (ODMR) in fluorescent NDs obtained by nanomilling and in the native bulk material. A wide-field fluorescence microscope was used to address the problem of a potentially heterogeneous response of NDs between and within samples.

Our results show markedly different dependencies on high-laser power for the ODMR contrast in bulk diamond and NDs. While contrast and $NV^-$ spin polarization steadily increase with laser power in the bulk diamond, with little to no variation across the sample, they show a more complex and heterogeneous behavior in the NDs.  In bare as-deposited NDs we observed a decrease in the ODMR contrast at the highest power, reflecting a more efficient $NV^- \rightarrow NV0$ photoconversion compared to the bulk, and implying a reduction in the pool of $NV^-$ centers. The non-monotonic behavior in NDs is likely to be determined by the interplay between spin and charge dynamics under continuous laser illumination. NDs showed a high tendency to aggregate, forming much more heterogeneous systems than in the homogeneous bulk diamond. For NDs internalized in cells we observed a qualitatively similar trend as in the bare NDs, with a different, more variable behavior at the highest laser power, suggesting that the cellular environment may have a role in the dynamics of NV charges, perhaps due to the different pH or to surface interactions with different proteins in the cytosol.

In conclusion, the large exposed surface area of NDs is greatly beneficial for sensing applications, e.g. in bioimaging assays, but the effects of surface states and surface interactions on the NV charge stability and photoconversion dynamics must be taken into account. Increasing laser power in the native bulk diamond increases ODMR contrast, a measure of the spin polarization of the ensemble of $NV^-$. Conversely, in NDs, surface effects may limit the benefits of stronger laser power due to photoconversion between different charge states. The effects reported here highlight a trade off in the use of NDs for sensing and polarization transfer applications.

**Acknowledgements**

We thank L. Basso and P. Aprà for Raman spectroscopy in the department of physics at University of Trento (Italy) and University of Torino (Italy), respectively.  We also thank Massimo Cazzanelli for the construction of the set-up and Prof. Simonetta Geninatti-Crich for her input and critical reading of the manuscript.

This project was funded by the European Union Horizon 2020 research and innovation program under the MSCA-ITN grant agreement 766402 (ZULF-NMR) and FET-OPEN grant agreement 858149 (AlternativesToGd). J.A. and A.K. acknowledge the funding by the Deutsche Forschungsgemeinschaft (grant KR3316/6-2, within the FOR1493).

24. Fischer, R.; Bretschneider, C.O.; London, P.; Budker, D.; Gershoni, D.; Frydman, L. Bulk nuclear polarization enhanced at room temperature by optical pumping. *Phys. Rev. Lett.* **2013**, *111*, 1–5, doi:10.1103/PhysRevLett.111.057601.

25. Frydman, L.; Fischer, R.; Onoda, S.; Isoya, J.; Álvarez, G.A.; Gershoni, D.; London, P.; Kanda, H.; Bretschneider, C.O. Local and bulk 13C hyperpolarization in nitrogen-vacancy-centred diamonds at variable fields and orientations. *Nat. Commun.* **2015**, *6*, 1–2, doi:10.1038/ncomms9456.

26. Ajoy, A.; Safvati, B.; Nazaryan, R.; Oon, J.T.; Han, B.; Raghavan, P.; Nirodi, R.; Aguilar, A.; Liu, K.; Cai, X.; et al. Hyperpolarized relaxometry based nuclear T 1 noise spectroscopy in diamond. *Nat. Commun.* **2019**, *10*, 1–16, doi:10.1038/s41467-019-13042-3.

27. Giri, R.; Gorrini, F.; Dorigoni, C.; Avalos, C.E.; Cazzanelli, M.; Tambalo, S.; Bifone, A. Coupled charge and spin dynamics in high-density ensembles of nitrogen-vacancy centers in diamond. *Phys. Rev. B* **2018**, *98*, 1–7, doi:10.1103/PhysRevB.98.045401.

28. Balasubramanian, P.; Osterkamp, C.; Chen, Y.; Chen, X.; Teraji, T.; Wu, E.; Naydenov, B.; Jelezko, F. Dc Magnetometry with Engineered Nitrogen-Vacancy Spin Ensembles in Diamond. *Nano Lett.* **2019**, *19*, 6681–6686, doi:10.1021/acs.nanolett.9b02993.

29. Kaviani, M.; Deák, P.; Aradi, B.; Frauenheim, T.; Chou, J.P.; Gali, A. Proper surface termination for luminescent near-surface NV centers in diamond. *Nano Lett.* **2014**, *14*, 4772–4777, doi:10.1021/nl501927y.

30. Nagl, A.; Hemelaar, S.R.; Schirhagl, R. Improving surface and defect center chemistry of fluorescent nanodiamonds for imaging purposes-a review. *Anal. Bioanal. Chem.* **2015**, *407*, 7521–7536, doi:10.1007/s00216-015-8849-1.

31. Krueger, A.; Lang, D. Functionality is key: Recent progress in the surface modification of nanodiamond. *Adv. Funct. Mater.* **2012**, *22*, 890–906, doi:10.1002/adfm.201102670.

32. McGuinness, L.P.; Yan, Y.; Stacey, A.; Simpson, D.A.; Hall, L.T.; Maclaurin, D.; Prawer, S.; Mulvaney, P.; Wrachtrup, J.; Caruso, F.; et al. Quantum measurement and orientation tracking of fluorescent nanodiamonds inside living cells. *Nat. Nanotechnol.* **2011**, *6*, 358–363, doi:10.1038/nnano.2011.64.

33. Wang, P.; Chen, S.; Guo, M.; Peng, S.; Wang, M.; Chen, M.; Ma, W.; Zhang, R.; Su, J.; Rong, X.; et al. Nanoscale magnetic imaging of ferritins in a single cell. *Sci. Adv.* **2019**, *5*, 1–6, doi:10.1126/sciadv.aau8038.

34. Parker, A.J.; Jeong, K.; Avalos, C.E.; Hausmann, B.J.M.; Vassiliou, C.C.; Pines, A.; King, J.P. Optically pumped dynamic nuclear hyperpolarization in C 13 -enriched diamond. *Phys. Rev. B* **2019**, *100*, doi:10.1103/PhysRevB.100.041203.

35. Boudou, J.P.; Tisler, J.; Reuter, R.; Thorel, A.; Curmi, P.A.; Jelezko, F.; Wrachtrup, J. Fluorescent nanodiamonds derived from HPHT with a size of less than 10 nm. *Diam. Relat. Mater.* **2013**, *37*, 80–86, doi:10.1016/j.diamond.2013.05.006.

36. Shenderova, O.; Nunn, N.; Oeckinghaus, T.; Torelli, M.; McGuire, G.; Smith, K.; Danilov, E.; Reuter, R.; Wrachtrup, J.; Shames, A.; et al. Commercial quantities of ultrasmall fluorescent nanodiamonds
18

# Supplementary Information

Divergent effects of laser irradiation on ensembles of nitrogen-vacancy centers in bulk and nano-diamonds: implications for biosensing

Domingo Olivares-Postigo[1,2,3]*, Federico Gorrini[2,4], Valeria Bitonto[3], Johannes Ackermann[5], Rakshyakar Giri[1], Anke Krueger[5,6] and Angelo Bifone[2,3,4]

[1] Center for Neuroscience and Cognitive Systems, Istituto Italiano di Tecnologia, Corso Bettini 31, 38068 Rovereto, Trento, Italy
[2] University of Torino, Molecular Biology Center, via Nizza 52, 10126, Torino, Italy
[3] University of Torino, Department of Molecular Biotechnology and Health Sciences, via Nizza 52, 10126, Torino, Italy
[4] Istituto Italiano di Tecnologia, Center for Sustainable Future Technologies, via Livorno 60, 10144, Torino, Italy
[5] Institut für Organische Chemie, Julius-Maximilians-Universität Würzburg, Am Hubland, 97074 Würzburg, Germany
[6] Wilhelm Conrad Röntgen Center for Complex Materials Research (RCCM), Julius-Maximilians University Würzburg 97074, Germany

* domingo.olivares@iit.it

**Table of contents**

## 1. SEM of milled fluorescent ND

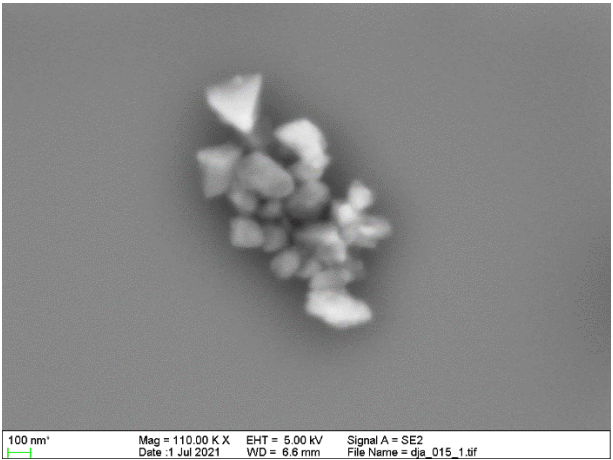
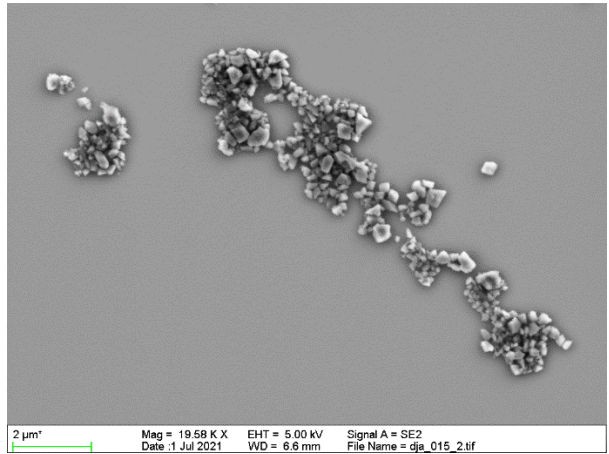
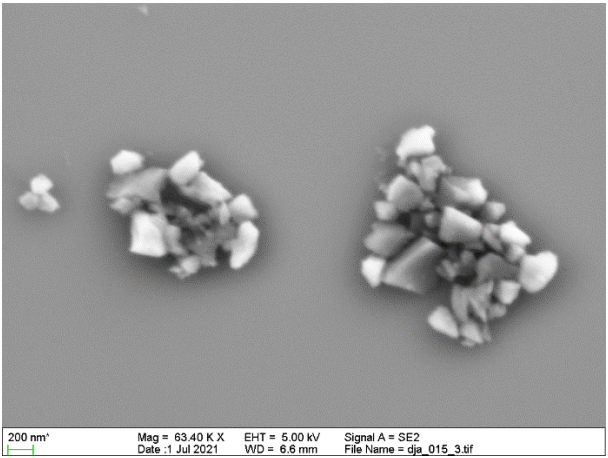

*Figure S1 |* *SEM images of milled fluorescent diamond (fND) before size separation by centrifugation and at different magnifications*

*The samples were prepared by drop-coating on silicon substrate with an aqueous solution of the milled fND.*

*Scale bars: 100 nm (upper left), 2 µm (upper right), 200 nm (lower left)*

## 2. Raman spectrum of milled fluorescent ND

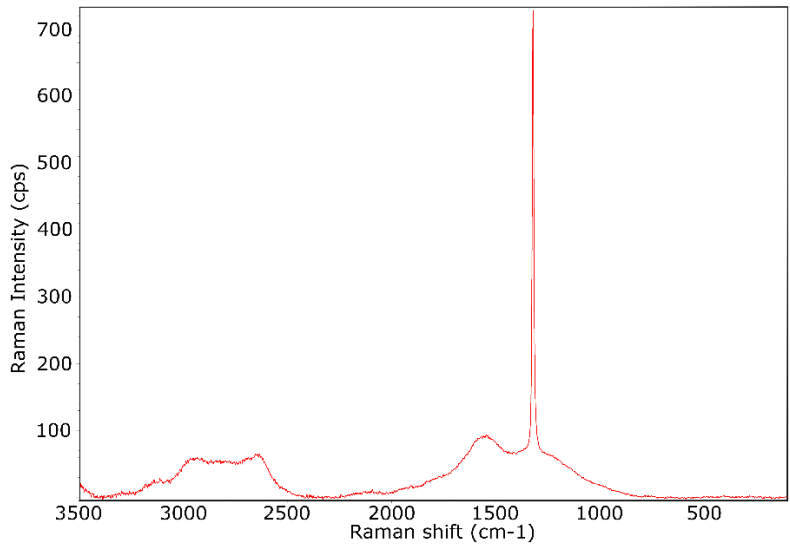

*Figure S2 | Raman spectrum of milled fND using laser excitation at a wavelength of 445 nm.*

## 3. Dynamic light scattering of aqueous colloids of milled fluorescent ND

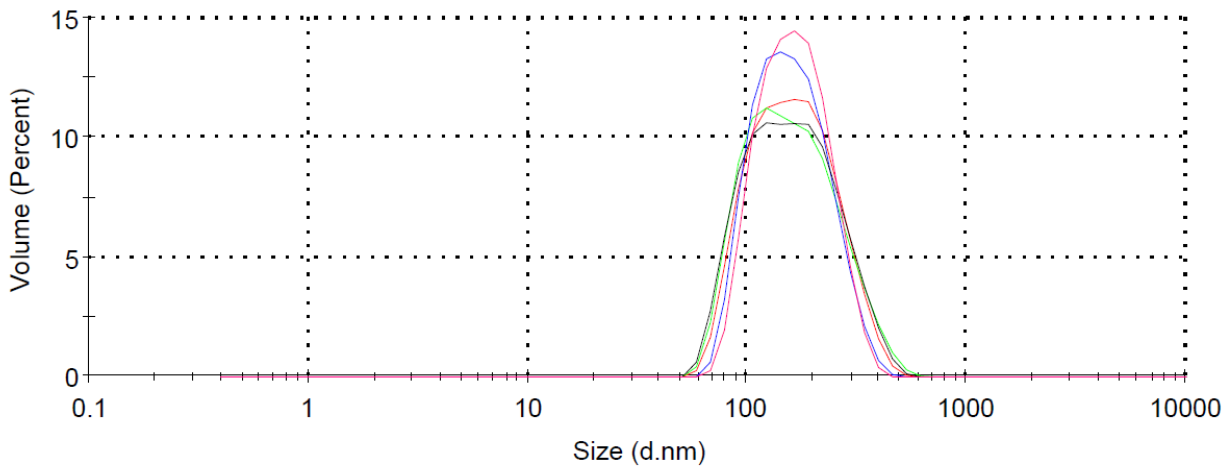

*Figure S3 | Dynamic light scattering of the colloidal fraction of milled fND centrifuged at 3000 rpm for 1 hour. The D50 value of the volume size distribution is 156 nm*

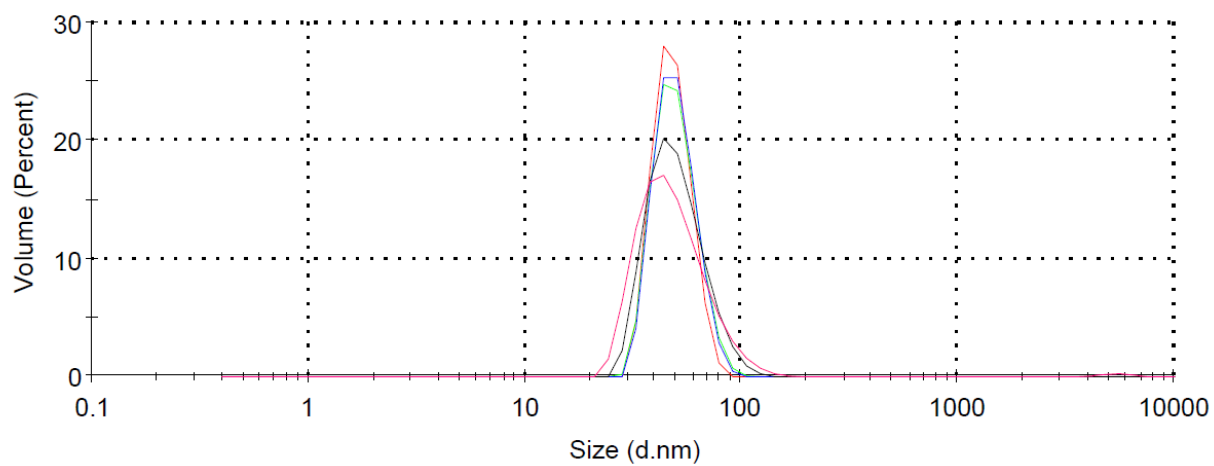

***Figure S4 |*** *Dynamic light scattering of the colloidal fraction of milled fND centrifuged at 15000 rpm for 1 hour. The D50 value of the volume size distribution is 48 nm.*